\documentclass[aps,12pt]{revtex4}
\usepackage{graphics}
\usepackage{graphicx}
\usepackage{amsmath}
\usepackage{amsfonts}
\usepackage{amssymb}

\begin{document}

\setlength{\unitlength}{1mm}

\title[TASEP with particle-dependent hopping probabilities on a ring]{Determinant solution for the TASEP with particle-dependent \\ hopping probabilities on a ring}

\author{V.S. Poghosyan}
\email{vpoghos@thsun1.jinr.ru}

\author{V.B. Priezzhev}
\email{priezzvb@thsun1.jinr.ru}

\affiliation{ Bogoliubov Laboratory of Theoretical Physics,
\\ Joint Institute for Nuclear Research,
\\ 141980 Dubna, Russia}

\bigskip

\begin{abstract}
We consider the totally asymmetric exclusion process on a ring in discrete time with the backward-ordered sequential
update and particle-dependent hopping probabilities.
Using a combinatorial treatment of the Bethe ansatz, we derive the determinant expression for the non-stationary
probability of transitions between particle configurations.
In the continuous-time limit, we find a generalization of the recent result, obtained by A. R\'akos and G.M. Sch\"utz for
infinite lattice, to the case of ring geometry.
\end{abstract}

\maketitle

\section{Introduction}

In recent paper \cite{RakosSchutz} R\'{a}kos and Sch\"{u}tz have solved the master equation for the one-dimensional
totally asymmetric exclusion process (TASEP) with particle-dependent rates defined on the integer lattice $\mathbb{Z}$.
The TASEP with particle-dependent rates \cite{Benjamini} generalizes the usual TASEP, where particles hop independently
in continuous time with equal rates to their right neighboring sites provided the target sites are empty
\cite{Liggett, SchutzBook}.
A random choice of rates leads to condensation phenomena which have been studied intensively for the associated zero range
processes \cite{EvansHanney,Grosskinsky,Godreche,Levine,Kaupuzs,Harris,GodrecheLuck}.
Specifically, the motion of the slowest particle generates a macroscopic traffic jam
viewed in the coarse-grained scale as a discontinuity in the particle density.
The solution of the master equation for the particle-dependent rates in \cite{RakosSchutz} was obtained under the general
approach developed by Sch\"{u}tz \cite{SchutzBook, Schutz} for the usual TASEP for a $P$-particle system on the infinite
one-dimensional lattice.
The conditional probability to find $P$ particles in positions $x_1,x_2,\ldots,x_P$; $x_1 < x_2 < \ldots < x_P$
at continuous time $t$ if these are in positions $x_1^0,x_2^0,\ldots,x_P^0$; $x_1^0 < x_2^0 < \ldots < x_P^0$ at initial
moment of time was found in the form of a determinant of $P\times P$ matrix.
In this paper, we extend the determinant formula for more complicated case, the TASEP in discrete time with backward sequential update
on the lattice with periodic boundary conditions.
The usual TASEP with the ring geometry has been considered in \cite{PriezPRL, PriezCondmat, PriezProceedings}.
Two methods, an analytical Bethe Ansatz and a combinatorial solution were presented.
Here, we use the combinatorial approach \cite{PriezCondmat}, based on a geometrical interpretation of the Bethe ansatz.

The discrete time TASEP with backward-sequential update is defined as follows.
At each discrete moment of time $T$, particles are tested for mobility successively in backward direction starting from an
arbitrary empty site.
A particle is movable at the moment $T$ if its right site is not occupied by next particle at time $T+1$.
(In the parallel update all particles are tested simultaneously and the particle is movable at time $T$ if its right site
is empty at time $T$).
Using a discrete time formulation, we interpret the Bethe ansatz as a combinatorial problem of cancellation of forbidden
space-time trajectories of particles.
A resulting expression for the conditional probability is then transformed into a continuous-time formula by changing the
one-particle discrete Bernoulli dynamics for the continuous Poisson process.

The paper organized as follows. The content of section II is a description of the combinatorial approach, which is
included for convenience and completeness. In subsection A of section II, we describe a combinatorial ansatz for
directed walks using as an example the free-fermion problem of "vicious" walkers \cite{Fisher}. In subsection B, we
introduce the combinatorial treatment of the Bethe ansatz and consider the discrete-time TASEP with same hopping
probability for all particles. The main subject of the paper, the TASEP with particle-dependent rates is considered in
section III. In section IV, we close the paper with some conclusions.

\section{The Combinatorial method}

\subsection{Combinatorial ansatz for directed walks}
Consider $P$ particles labelled $\{1,2,\ldots,P\}$ hopping in one direction
on an infinite one-dimensional lattice. The discrete space-time dynamics of
their motion can be described as a set of continuous broken trajectories on
a triangle lattice $\Lambda$, which is obtained from the square lattice by
adding a diagonal bond between the upper left corner and the lower right
corner of each elementary square. Let $(x,T)$ be integer space-time coordinates
of a particle on $\Lambda$, where the vertical time axis is directed down and
the horizontal space axis is directed right. A trajectory of each particle is
a sequence of connected vertical and diagonal bonds of $\Lambda$. The diagonal
bonds correspond to jumps of particles to their right for
a unit time-step with probability $v$. The vertical bonds correspond to stays
of particles at fixed sites during the unit time-step with probability
$1-v$. The initial positions of particles are $x^0_1 < x^0_2 < \ldots < x^0_P$.
The problem of "vicious" walkers \cite{Fisher} is to find the conditional
probability $P_T(\vec{x}|\vec{x}^0)$ for the particles to reach the positions
$\vec{x}=(x_1,x_2,\ldots,x_P)$; $x_1 < x_2 < \ldots < x_P$
from the initial positions $\vec{x}^0=(x^0_1,x^0_2,\ldots,x^0_P)$ for time $T$ so that no pairs of trajectories
are intersected during time $T$. It means that every site can be occupied by at most one particle
at every moment of discrete time and if this rule is violated at a moment $T'<T$, the process stops.

We start with consideration of one-particle motion on the infinite lattice.
Let $\mathcal{T}_T(x|x^0)$ be a set of one-particle trajectories, which are
starting at $(x^0,0)$ and finishing at $(x,T)$. Each trajectory
$q\in \mathcal{T}_T(x|x^0)$ is realized with probability
$v^{x-x^0}(1-v)^{T-x+x^0}$. Therefore, the total
probability for the particle to reach $x$ from $x^0$ for time $T$ is
\begin{equation}
P_T(x|x^0)=v^{x-x^0}(1-v)^{T-x+x^0}\|\mathcal{T}_T(x|x^0)\|=
F_0(x-x^0|T).
\label{OneParticleProb}
\end{equation}
where
\begin{equation}
F_0(x|T)=\binom{T}{x}v^x(1-v)^{T-x}
\label{F0}
\end{equation}
is the statistical weight of the set $\mathcal{T}_T(x|x^0)$ of one-particle trajectories.
For the case of $P$ particles, the set ${\mathbb S}_P$ of all possible
free trajectories is a direct product of
one-particle sets of trajectories reaching $(\vec{x},T)$ from $(\vec{x}^0,0)$:
\begin{equation}
{\mathbb S}_P=
\mathcal{T}_T(x_1|x_1^0)\otimes
\mathcal{T}_T(x_2|x_2^0)\otimes\ldots\otimes
\mathcal{T}_T(x_P|x_P^0)
\label{SP}
\end{equation}
The set ${\mathbb S}_P$ contains non-intersecting and intersecting trajectories.
The latter cases are unallowed by the single-occupation rule and should be subtracted in evaluations
of probability $P_T(\vec{x}|\vec{x}^0)$.

Consider first the case $P=2$. We denote the set of unallowed elements by $\mathbb{U}_2\subset{\mathbb S}_2$.
To cancel the contribution of forbidden elements, we introduce an auxiliary set of pairs of trajectories
\begin{equation}
{\mathbb A}_{21}=\mathcal{T}_T(x_2|x_1^0)\otimes\mathcal{T}_T(x_1|x_2^0),
\end{equation}
where the final coordinates are permuted with respect to those in ${\mathbb S}_2$.
It is easy to see that all elements of the set ${\mathbb A}_{21}$ are pairs of
crossing trajectories.
Each intersecting pair $q \in {\mathbb S}_2$ has a first collision point $(x_c,T_c)$, i.e.
the space-time point, where the trajectories meet for the first time.
Consider the set ${\mathbb S}_2(x_c,T_c)$ of intersecting trajectories with the fixed
first collision point $(x_c,T_c)$.
The sets ${\mathbb S}_2(x_c,T_c),\;(x_c,T_c)\in \Lambda$ break the set $\mathbb{U}_2$
into subsets parameterized by coordinates $(x_c,T_c)$.
For all $(x_c',T_c')\neq (x_c'',T_c'')$ we have
\begin{equation}
{\mathbb S}_2(x_c',T_c')\cap {\mathbb S}_2(x_c'',T_c'')=\emptyset
\label{emptyS}
\end{equation}
and
\begin{equation}
\bigcup_{(x_c,T_c)\in \Lambda} {\mathbb S}_2(x_c,T_c) = \mathbb{U}_2
\label{U2}
\end{equation}

For every set ${\mathbb S}_2(x_c,T_c)$, there exists a uniquely defined set
${\mathbb A}_{21}(x_c,T_c)\subset {\mathbb A}_{21}$ obtained from ${\mathbb S}_2(x_c,T_c)$ by
permutation of tails of all trajectories beginning at the first collision point.
The sets ${\mathbb S}_2(x_c,T_c)$ and ${\mathbb A}_{21}(x_c,T_c)$ are geometrically
equivalent, because there is one-to-one correspondence between their elements.
Therefore $\|{\mathbb S}_2(x_c,T_c)\|=\|{\mathbb A}_{21}(x_c,T_c)\|$.
Like the sets ${\mathbb S}_2(x_c,T_c)$, the sets
${\mathbb A}_{21}(x_c,T_c)$ break the set ${\mathbb A}_{21}$ into subsets: for all $(x_c',T_c')\neq (x_c'',T_c'')$ we have
\begin{equation}
{\mathbb A}_{21}(x_c',T_c')\cap {\mathbb A}_{21}(x_c'',T_c'')=\emptyset
\label{emptyA}
\end{equation}
and
\begin{equation}
\bigcup_{(x_c,T_c)\in \Lambda} {\mathbb A}_{21}(x_c,T_c) = {\mathbb A}_{21}.
\label{A21}
\end{equation}

It follows from Eqs.(\ref{emptyS},\ref{U2},\ref{emptyA},\ref{A21}) that the whole sets $\mathbb{U}_2$
and ${\mathbb A}_{21}$ are geometrically equivalent and
$\|\mathbb{U}_2\|=\|{\mathbb A}_{21}\|$.
Given $(\vec{x},T)$ and $(\vec{x}^0,0)$, all elements of ${\mathbb S}_{2}$ have the same weight
\begin{eqnarray}
Q=\prod_{i=1}^{2}v^{x_i-x_i^0}(1-v)^{T-x_i+x_i^0}.
\end{eqnarray}
Then, we have for the probability
\begin{eqnarray}
P_T(\vec{x}|\vec{x}^0)&=&Q \times \left(\|{\mathbb S}_2\|-\|{\mathbb A}_{21}\|\right)=\\
\nonumber
&=& \left( F_0(x_1-x_1^0|T)F_0(x_2-x_2^0|T)-F_0(x_2-x_1^0|T)F_0(x_1-x_2^0|T)\right)=\\ \nonumber
&=&det\; \mathbf{M},
\end{eqnarray}
where the elements of the $2\times 2$ matrix $\mathbf{M}$ are
\begin{equation}
M_{i,j}=F_0(x_j-x_i^0|T),\hspace{.5cm} i,j=1,2.
\label{M2}
\end{equation}

This result can be easily generalized for the $P$-particle system of vicious walkers \cite{Fisher}.
The set ${\mathbb S}_P$ contains the subset of intersecting trajectories ${\mathbb U}_P$.
We introduce $P!-1$ auxiliary sets
\begin{equation}
{\mathbb A}_{\pi(12\ldots P)}=
\mathcal{T}_T( x_{\pi(1)} |x_1^0)\otimes
\mathcal{T}_T( x_{\pi(2)} |x_2^0)\otimes\ldots\otimes
\mathcal{T}_T( x_{\pi(P)} |x_P^0)
\end{equation}
where $\pi$ is any permutation of numbers $1,2,\ldots,P$ beside the identical one $\hat{1}$.
Each element of the set
\begin{equation}
{\mathbb U}_P\bigcup\left(\bigcup_{\pi\neq \hat{1}} {\mathbb A}_{\pi(1,2,\ldots,P)}\right)
\label{UA}
\end{equation}
containing trajectories intersecting at given point $(x_c,T_c)$,
has a unique geometrically identical counterpart
with a pair of permuted indices, e.g. $1,2,\ldots,i,\ldots,j,\ldots,P$ and $1,2,\ldots,j,\ldots,i,\ldots,P$. Introducing
the sign of permutation $sgn(\pi)$, we can ascribe opposite signs to two counterparts.
Then, we obtain
\begin{eqnarray}
P_T(\vec{x}|\vec{x}^0)&=&Q \times \left(\|{\mathbb S}_P\|+
\sum_{\pi \neq \hat{1}}sgn(\pi)\|{\mathbb A}_{\pi(12\ldots P)}\|\right)=\\
\nonumber
&=&det\; \mathbf{M},
\label{det1}
\end{eqnarray}
where
\begin{equation}
M_{i,j}=F_0(x_j-x_i^0|T),\hspace{.5cm} i,j=1,2,\ldots P
\label{Mij}
\end{equation}
and
\begin{eqnarray}
Q=\prod_{i=1}^{P}v^{x_i-x_i^0}(1-v)^{T-x_i+x_i^0}.
\label{QPF}
\end{eqnarray}

\subsection{Combinatorial treatment of Bethe ansatz}

The vicious walkers are locally interacting particles. In this subsection,
we consider the system of particles with the interaction of a non-zero range,
which takes place in the case of the totally asymmetric exclusion process with
the backward sequential update \cite{Schreck1,Schreck2}.
Consider again $P$ particles hopping in one direction on an infinite one-dimensional lattice.
The interaction between particles can be defined by the following rules:

1. Trajectories of particles do not intersect (every site can be occupied
by at most one particle.

2. A particle stays at its own site with probability $1$ if the
target site is occupied by another particle during the step of discrete time.

Like the previous case, the trajectory of each particle is a sequence of
vertical and diagonal bonds on the lattice $\Lambda$.
Each diagonal bond has weight $v$ and the vertical bond weight $1-v$.
In view of more complicated interaction, it is convenient to decompose the set of all
free trajectories into more detailed subsets.
For each vertical bond $[(x,t),(x,t+1)]$, the trajectory of $i$-th particle passing this
bond can be decomposed into two parts
$[(x_i^0,0) \to (x,t)|1|(x,t+1) \to (x_i,T)]$ and $[(x_i^0,0) \to (x,t)|-v|(x,t+1) \to (x_i,T)]$
where the value between vertical bars means the weight of the bond $[(x,t),(x,t+1)]$.
These new trajectories are geometrically equivalent, but the first trajectory passes
the selected vertical bond with weight $1$ and second one with weight $-v$.
We make this decomposition for each vertical bond of each trajectory and denote the
whole set of decomposed trajectories of $i$-th particle by $\mathcal{T}_T(x_i|x_i^0)$
using the same notation as for the set of one-particle trajectories in the previous
subsection.
The weight of the trajectory $q \in \mathcal{T}_T(x_i|x_i^0)$ is a product of the
weights of its bonds.
\begin{equation}
\mu(q)=\prod_{k=1}^{T}\mu(k\texttt{\small -th bond of }q)
\end{equation}
The weight of the set of trajectories is a sum of weights of its elements.
In accordance with the above definition (\ref{OneParticleProb}), the weight of the whole set
$\mathcal{T}_T(x_i|x_i^0)$ is
\begin{equation}
\mu(\mathcal{T}_T(x_i|x_i^0))= F_0(x_i-x_i^0|T)
\end{equation}
The set of free $P$-particle trajectories ${\mathbb S}_P$ is defined by Eq. (\ref{SP}) and has the weight
\begin{equation}
\mu({\mathbb S}_P)=\sum_{q\in {\mathbb S}_P}\mu(q)=
\prod_{i=1}^{P}v^{x_i-x^0_i}(1-v)^{T-x_i + x_i^0} \binom{T}{x_i-x^0_i}.
\end{equation}
As in the case of vicious walkers, the set ${\mathbb S}_P$ contains a subset of
unallowed elements,${\mathbb U}_P$, which should be excluded. By the rules 1 and 2,
an element of ${\mathbb S}_P$ is unallowed, if there is at least one
pair of intersecting trajectories, or if there are two neighboring vertical bonds from which
the left one has weight $-v$.
To cancel unallowed elements we start as above with the case $P=2$ and construct
an auxiliary set of trajectories ${\mathbb A}_{21}$. We will show that, due to non-locality, the
auxiliary set has more complicated structure

\begin{equation}
{\mathbb A}_{21}=\bigcup_{k_1=0}^{\infty}\bigcup_{k_2=0}^{1}
\biggl(
\mathcal{T}_T(x_2|x_1^0-k_1) \otimes \mathcal{T}_T(x_1|x_2^0-k_2)
\biggr) .
\label{defA21}
\end{equation}

Beginning the construction, we notice that every unallowed pair has
a first collision point $(x_c,T_c)\in \Lambda$,
where the particles come for the first time to neighboring sites at a moment $T_c<T$.
The first trajectory reaches site $(x_c,T_c)$ from $(x_1^0,0)$ and the second one reaches site $(x_c+1,T_c)$
from $(x_2^0,0)$ and then it makes the vertical step to site $(x_c+1,T_c+1)$.

The first trajectory has, just after the collision, a diagonal bond $[(x_c,T_c),(x_c+1,T_c+1)]$
with weight $v$ (referred to as a collision of the first type) or
vertical bond $[(x_c,T_c),(x_c,T_c+1)]$ with weight $-v$ (refereed to as the
collision of the second type).
Notice, that trajectories, which have the vertical bond $[(x_c,T_c),(x_c,T_c+1)]$ with
weight $1$ are allowed.
Thus, we have two types of unallowed elements of ${\mathbb U}_2$ for fixed $(x_c,T_c)$.
Denote these subsets by ${\mathbb V}(x_c,T_c)$ for the first type and
$ {\mathbb W}(x_c,T_c)$ for the second one.
For all $(x_c',T_c')\neq (x_c'',T_c'')$ we have
\begin{eqnarray}
{\mathbb V}(x_c',T_c')&\cap& {\mathbb V}(x_c'',T_c'')=\emptyset,\nonumber\\
{\mathbb W}(x_c',T_c')&\cap& {\mathbb W}(x_c'',T_c'')=\emptyset,\nonumber\\
{\mathbb V}(x_c',T_c')&\cap& {\mathbb W}(x_c'',T_c'')=\emptyset,\nonumber
\end{eqnarray}
and
\begin{equation}
\bigcup_{(x_c,T_c)\in \Lambda}
\biggl(
{\mathbb V}(x_c,T_c) \cup {\mathbb W}(x_c,T_c)
\biggr)
=\mathbb{U}_2
\label{u2}
\end{equation}

For each space-time point $(x_c,T_c)\in \Lambda$, we construct the sequences of pairs of trajectories
${\mathbb A}_{v}(k_1,k_2)$ and ${\mathbb A}_{w}(k_1,k_2)$ with $k_1=0,1,2,\ldots$ and $k_2=0,1$
(arguments $x_c$ and $T_c$ are omitted), obtained from ${\mathbb V}(x_c,T_c)$ and ${\mathbb W}(x_c,T_c)$ by permutation
of tails in each pair of trajectories and shifting the initial parts in negative direction.
Explicitly, the transformation ${\mathbb V}(x_c,T_c) \Rightarrow {\mathbb A}_{v}(k_1,k_2)$ is
\begin{equation}
\begin{pmatrix}
(x_1^0,0)\\
\downarrow\\
(x_c,T_c)\\
\downarrow\\
(x_c+1,T_c+1)\\
\downarrow\\
(x_1,T)
\end{pmatrix}
\otimes
\begin{pmatrix}
(x_2^0,0)\\
\downarrow\\
(x_c+1,T_c)\\
\downarrow\\
(x_c+1,T_c+1)\\
\downarrow\\
(x_2,T)
\end{pmatrix}
\Rightarrow
\begin{pmatrix}
(x_1^0-k_1,0)\\
\downarrow\\
(x_c-k_1,T_c)\\
\downarrow\\
(x_c+1-k_1,T_c+1)\\
\downarrow\\
(x_2,T)
\end{pmatrix}
\otimes
\begin{pmatrix}
(x_2^0-k_2,0)\\
\downarrow\\
(x_c+1-k_2,T_c)\\
\downarrow\\
(x_c+1-k_2,T_c+1)\\
\downarrow\\
(x_1,T)
\end{pmatrix}
\nonumber
\end{equation}
and the transformation ${\mathbb W}(x_c,T_c) \Rightarrow {\mathbb A}_{w}(k_1,k_2)$ is
\begin{equation}
\begin{pmatrix}
(x_1^0,0)\\
\downarrow\\
(x_c,T_c)\\
\downarrow\\
(x_c,T_c+1)\\
\downarrow\\
(x_1,T)
\end{pmatrix}
\otimes
\begin{pmatrix}
(x_2^0,0)\\
\downarrow\\
(x_c+1,T_c)\\
\downarrow\\
(x_c+1,T_c+1)\\
\downarrow\\
(x_2,T)
\end{pmatrix}
\Rightarrow
\begin{pmatrix}
(x_1^0-k_1,0)\\
\downarrow\\
(x_c-k_1,T_c)\\
\downarrow\\
(x_c-k_1,T_c+1)\\
\downarrow\\
(x_2,T)
\end{pmatrix}
\otimes
\begin{pmatrix}
(x_2^0-k_2,0)\\
\downarrow\\
(x_c+1-k_2,T_c)\\
\downarrow\\
(x_c+1-k_2,T_c+1)\\
\downarrow\\
(x_1,T)
\end{pmatrix}
\nonumber
\end{equation}
It follows from the definitions that
\begin{equation}
\bigcup_{(x_c,T_c)\in \Lambda}\bigcup_{k_1=0}^{\infty}\bigcup_{k_2=0}^{1}
\biggl(
{\mathbb A}_{v}(k_1,k_2)\cup{\mathbb A}_{w}(k_1,k_2)
\biggr) = {\mathbb A}_{21}.
\label{normA21}
\end{equation}
The weights of introduced sets obey the following relations
\begin{eqnarray}
\mu\left({\mathbb A}_{v}(0,0)\right)&= &\mu\left({\mathbb V}(x_c,T_c)\right),\\
\mu\left({\mathbb A}_{v}(0,1)\right)&=-&\mu\left({\mathbb W}(x_c,T_c)\right),\\
\mu\left({\mathbb A}_{w}(k,0)\right)&=-&\mu\left({\mathbb A}_{v}(k+1,0)\right),\;k=0,1,2\ldots\;,\\
\mu\left({\mathbb A}_{w}(k,1)\right)&=-&\mu\left({\mathbb A}_{v}(k+1,1)\right),\;k=0,1,2\ldots\;.
\end{eqnarray}
Using these relations, we can  express the probability
\begin{equation}
P_T(\vec{x}|\vec{x}^0)= \mu({\mathbb S}_2\setminus{\mathbb U}_2)
\label{PT2}
\end{equation}
via the weight of all unallowed configurations
\begin{equation}
\mu({\mathbb U}_2)=
\sum_{(x_c,T_c)}\sum_{k=0}^{\infty}
\biggl(
\mu({\mathbb A}_{v}(k,0))+\mu({\mathbb A}_{w}(k,0))-
\mu({\mathbb A}_{v}(k,1))-\mu({\mathbb A}_{w}(k,1))
\biggr).
\label{muU2}
\end{equation}

Our next aim is to bring this expression to a determinant form similar to (\ref{det1}).
Consider operator $\hat{a}_i$, which acts on the set of the free trajectories
of $i$-th particle and gives the set of free trajectories with the origin shifted by $1$ in negative direction:
\begin{equation}
\hat{a}_i \mathcal{T}_T(x_j|x_i^0)=\mathcal{T}_T(x_j|x_i^0-1)
\end{equation}
The operator representation allows us to write (\ref{muU2}) in the compact form:
\begin{equation}
\mu({\mathbb U}_2)=\mu\left(\frac{1-\hat{a}_2}{1-\hat{a}_1}\mathcal{T}_T(x_2|x_1^0)
\otimes\mathcal{T}_T(x_1|x_2^0)\right),
\label{oper}
\end{equation}
where the denominator is defined by its expansion
\begin{equation}
\frac{1}{1-\hat{a}_1}=\sum_{k=0}^{\infty}\hat{a}_1^k.
\end{equation}
and the summation of operators means the joining of the sets.

The crucial property of the operator expression in Eq.(\ref{oper}) is its factorization
with respect to indices 1 and 2. Introducing the functions
\begin{equation}
F_1(x_2-x_1^0|T)=\mu\left((1-\hat{a}_1)^{-1}\mathcal{T}_T(x_2|x_1^0)\right)=
\sum_{k=0}^{\infty}F_0(x_2-x_1^0+k|T),
\label{F1}
\end{equation}
and
\begin{equation}
F_{-1}(x_1-x_2^0|T)=\mu\left((1-\hat{a}_2)\mathcal{T}_T(x_1|x_2^0)\right)=
\sum_{k=0}^{1}(-1)^k F_0(x_1-x_2^0+k|T)
\label{F-1}
\end{equation}
we obtain the two-particle probability
$P_T(\vec{x}|\vec{x}^0)= \mu({\mathbb S}_2)-\mu({\mathbb U}_2) $
in the form
\begin{equation}
P_T(\vec{x}|\vec{x}^0)=F_{0}(x_1-x_1^0|T)F_{0}(x_2-x_2^0|T)
-F_{1}(x_2-x_1^0|T)F_{-1}(x_1-x_2^0|T)
\label{prob2}
\end{equation}
or
\begin{equation}
P_T(\vec{x}|\vec{x}^0)= det\; \mathbf{M},
\label{det2}
\end{equation}
where
\begin{equation}
M_{i,j}=F_{i-j}(x_i-x_j^0|T)\hspace{1cm}i,j=1,2
\label{M12}
\end{equation}

To generalize the determinant formula (\ref{det2}) to the general P-particle case, we should take into consideration
several additional properties of intersecting trajectories. First, we may organize the procedure of exclusion
of unallowed elements of ${\mathbb U}_P$ in an ordered way. Starting with the top line of the lattice $\Lambda$, we examine
all free trajectories of the set ${\mathbb S}_{P}$, row by row, until we meet the first collision point where unallowed
trajectories are cancelled with elements of the auxiliary set ${\mathbb A}_{\pi(12\ldots P)}$.

If the number of particles $P > 2$, the elementary squares associated with the collision points $(x_c,T_c)$,
$[(x_c,T_c),(x_c+1,T_c),(x_c,T_c+1),(x_c+1,T_c+1)]$ for different pairs of trajectories
may occur several times in one horizontal strip of $\Lambda$. If  squares
filled by interacting trajectories are separated one from another by a gap of
empty sites, the above arguments can be applied to each pair of interacting
trajectories independently. The crucial case for the Bethe ansatz is a situation, when the elementary squares are nearest
neighbors.
The specific property of the totally ASEP is that, in each pair of interacting trajectories, the right
trajectory remains free and interacts with the next trajectory independently
on its left neighbors. Therefore, we can analyse the interaction between
particles considering successively elementary squares in each row from left to
right  starting from an arbitrary empty square until all unwanted trajectories
will be removed.

After $T'$ steps from the top to bottom, one obtains the set of path configurations
$q \in {\mathbb S}_{P}$, which are allowed in the first $T'$ rows, and the set of
auxiliary configurations yet not involved into cancellation procedures.
Remembering that all elements of ${\mathbb A}_{\pi(12\ldots P)}$ for all $\pi\neq \hat{1}$
have end points permuted with respect to the original order $1,2,\ldots,P$, we
conclude that each element of ${\mathbb A}_{\pi(12\ldots P)}$ contains at least
one collision point. Therefore, all elements of the auxiliary set will be
cancelled after $T'=T$ steps with unallowed elements of ${\mathbb U}_P$.

The described way of exclusion of unallowed configurations implies a successive
construction of the auxiliary set ${\mathbb A}_{\pi(12\ldots P)}$. We notice that each
two intersecting trajectories in ${\mathbb A}_{21}$ are non-equivalent:
one of them belongs to the particle which overtakes another and can be called "active".
On the contrary, the second particle is "passive".
In the case $P>2$, one trajectory can overtake $m$ others, and we call it "m-active".
Similarly, the "m-passive" trajectories appear.

Assume, that the trajectory of a given particle has $m$ active intersections.
It means that it participates $m$ times in the cancellation procedure and its
starting point is shifted $m$ times to arbitrary distances in the negative
direction.
As a result, the auxiliary set associated with the free trajectory between points $x_i^0$ and $x_j$ becomes
\begin{equation}
\frac{1}{(1-\hat{a}_i)^m}\mathcal{T}_T(x_j|x_i^0)
\end{equation}
Similarly, for trajectories having $m$ passive intersections we get
\begin{equation}
(1-\hat{a}_i)^m\mathcal{T}_T(x_j|x_i^0)
\end{equation}
Note that these expressions may be combined into one, if we assume that $m>0$ for active trajectories and $m<0$ for
passive and $m=0$ for free ones.
The weights of sets of m-active and m-passive trajectories are given by functions introduced in \cite{PriezCondmat}:
\begin{equation}
F_m(x_j-x_i^0|T)=\sum_{k=0}^{\infty}\binom{k+m-1}{m-1} F_0(x_j-x_i^0+k|T)
\label{Fplus}
\end{equation}
for $m>0$, and
\begin{equation}
F_m(x_j-x_i^0|T)=\sum_{k=0}^{-m}(-1)^k\binom{-m}{k} F_0(x_j-x_i^0+k|T)
\label{Fminus}
\end{equation}
for $m<0$.
Activity $m$ of each trajectory is defined uniquely by the permutation $\pi(12\ldots P)$, so we have for the weight of
auxiliary set ${\mathbb A}_{\pi(12\ldots P)}$
\begin{equation}
\mu({\mathbb A}_{\pi(12\ldots P)})= \prod_{i=1}^{P}F_{\pi(i)-i}(x_{\pi(i)}-x_i^0|T)
\end{equation}
This product together with $sgn(\pi)$ is a term of expansion of the determinant $ det\; \mathbf{M}$ with matrix elements for any permutation
differing from the identical one
\begin{equation}
M_{i,j}=F_{i-j}(x_i-x_j^0|T)\hspace{1cm}i,j=1,2,\ldots,P
\label{M1P}
\end{equation}
It follows from Eq. (\ref{SP}) that the term corresponding to the identical
permutation $\pi=\hat{1}$ is
\begin{equation}
\mu({\mathbb S}_{P})= \prod_{i=1}^{P}F_{0}(x_{i}-x_i^0|T)
\end{equation}
Collecting the contributions from the set ${\mathbb S}_{P}$ and all
auxiliary sets ${\mathbb A}_{\pi(12\ldots P)}$, we obtain the the determinant
formula
\begin{equation}
P_T(\vec{x}|\vec{x}^0)= det\; \mathbf{M},
\label{detP}
\end{equation}
which is valid for all $P\geq 1$.
The TASEP with equal hopping probabilities on the ring has been considered in \cite{PriezCondmat}.
We skip the details of this derivation and consider ring geometry just for the TASEP with particle-dependent hopping probabilities.

\section{The Totally Asymmetric Exclusion process with particle-dependent hopping probabilities}

A method of solution of the TASEP with particle-dependent hopping probabilities follows in many details that of the previous section.
As above, we start with the case of two particles $P=2$ on the infinite lattice. The conditional probability (\ref{PT2})
is expressed as the weight of all allowed subsets $\mu({\mathbb S}_2\setminus{\mathbb U}_2)$, which depends on the hopping
probabilities of two particles $v_1$, $v_2$. The set of all possible free trajectories has the weight
\begin{equation}
\mu({\mathbb S}_2)=Q\times\|{\mathbb S}_2\|,
\end{equation}
where
\begin{eqnarray}
Q=\prod_{i=1}^{2}v_i^{x_i-x_i^0}(1-v_i)^{T-x_i+x_i^0}.
\label{Q2}
\end{eqnarray}
The weight of unallowed set $\mu({\mathbb U}_2)$ is determined by the structure of subsets
${\mathbb V}(x_c,T_c) \subset {\mathbb U}_2$, ${\mathbb W}(x_c,T_c) \subset {\mathbb U}_2$ and auxiliary sets
${\mathbb A}_{v}(k_1,k_2) \subset {\mathbb A}_{21}$ and ${\mathbb A}_{w}(k_1,k_2) \subset {\mathbb A}_{21}$ defined by
Eqs. (\ref{u2}), (\ref{normA21}) (see also the text between these expressions).
The idea of evaluation of $\mu({\mathbb U}_2)$ is to extract the factor $Q$ from the weight of ${\mathbb U}_2$, reducing
the problem of evaluation of $\mu({\mathbb U}_2)$ to consideration of the weights of bonds at the collision point
$(x_c,T_c)$ only.

\begin{figure}[h!]
\includegraphics[width=100mm]{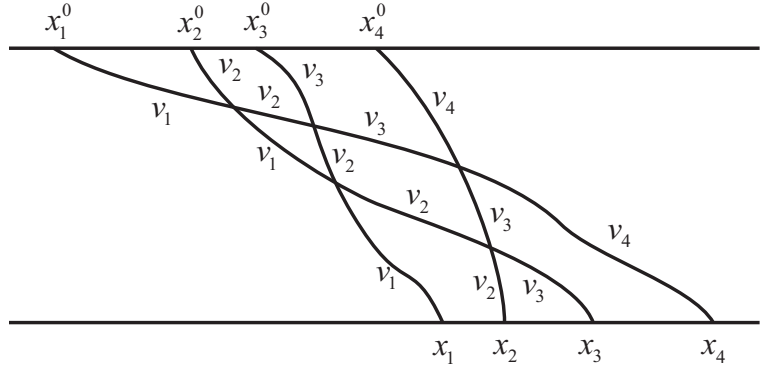}
\caption{\label{fig1} The set of four trajectories, which are active or passive in different intersections.}
\end{figure}

Extracting $Q$ from the weights of free trajectories, we get unweighted skeletons of trajectories, with fixed starting
and ending points.
Thus, we may consider a cancellation of skeletons of unallowed sets of trajectories instead of the cancellation of weighted
ones.
This is true also for subsets ${\mathbb V}(x_c,T_c)$ and ${\mathbb A}_v(k_1,k_2)$ as they consist of free trajectories
weighted in a standard way: the step of $i$-th particle has weight $v_i$, the stay of $i$-th particle has weight $(1-v_i)$.
However, the step $[(x_c,T),(x_c,T+1)]$ in ${\mathbb W}(x_c,T_c)$ and ${\mathbb A}_w(0,0)$ having originally the weight
$-v_i$, becomes $-\frac{v_i}{1-v_i}$ after extraction of $Q$, which leads to a correction factor in the total weights of
sets ${\mathbb W}(x_c,T_c)$ and ${\mathbb A}_{w}(0,0)$.
Therefore, the reduced weights of sets remaining after extraction can be written as
\begin{eqnarray}
\tilde{\mu}\left({\mathbb S}_{2}\right)     &=&\|{\mathbb S}_{2}\|\\
\tilde{\mu}\left({\mathbb V}(x_c,T_c)\right)&=&\|{\mathbb V}(x_c,T_c)\|\\
\tilde{\mu}\left({\mathbb A}_{v}(0,0)\right)&=&\|{\mathbb A}_{v}(0,0)\|\\
\tilde{\mu}\left({\mathbb W}(x_c,T_c)\right)&=&-\left(\frac{v_i}{1-v_i}\right)\|{\mathbb W}(x_c,T_c)\|\\
\tilde{\mu}\left({\mathbb A}_{w}(0,0)\right)&=&-\left(\frac{v_i}{1-v_i}\right)\|{\mathbb A}_{w}(0,0)\|,
\end{eqnarray}
where $\tilde{\mu}\left(\cdot\right)$ is defined by the identity $\mu \left(\cdot\right)=Q \,\tilde{\mu}\left(\cdot\right)$.
Along with reduced weights of ${\mathbb A}_{v}(0,0)$ and ${\mathbb A}_{w}(0,0)$, we define
$\tilde{\mu}\left({\mathbb A}_{v}(k_1,k_2)\right)$ and $\tilde{\mu}\left({\mathbb A}_{w}(k_1,k_2)\right)$ for
$k_1=0,1,2\ldots$ and $k_1=0,1$:
\begin{eqnarray}
\tilde{\mu}\left({\mathbb A}_{v}(0,1)\right) &=& \tilde{\mu}\left({\mathbb W}(x_c,T_c)\right),\\
-\left(\frac{v_i}{1-v_i}\right)\tilde{\mu}\left({\mathbb A}_{w}(k,0)\right) &=&
                               \tilde{\mu}\left({\mathbb A}_{v}(k+1,0)\right),\;k=0,1,2\ldots\;,\\
-\left(\frac{v_i}{1-v_i}\right)\tilde{\mu}\left({\mathbb A}_{w}(k,1)\right) &=&
                               \tilde{\mu}\left({\mathbb A}_{v}(k+1,1)\right),\;k=0,1,2\ldots\;.
\end{eqnarray}
Using these relations, we can write the weight of unallowed set in an operator form and verify the identity
\begin{equation}
\mu({\mathbb U}_2)=Q\times \tilde{\mu}
\left(
\frac{1-\frac{v_1}{1-v_1}\hat{a}_2}
     {1-\frac{v_1}{1-v_1}\hat{a}_1}\;
\mathcal{T}_T(x_2|x_1^0)\otimes
\mathcal{T}_T(x_1|x_2^0)
\right)
\label{muU2_2}
\end{equation}
An important property of Eq. (\ref{muU2_2}) is that the reduced weight of two interacting trajectories depends on the weight
of only one of them, namely on that named "active" in the previous section. Therefore, given a permutation of $P$ end points we can
evaluate easily the reduced weight of each trajectory taking into account its all active and passive intersections with other trajectories.
Another important property of Eq. (\ref{muU2_2}) is that the operators $\hat{a}_1$ and $\hat{a}_2$ enter in a factorized form.
This property guarantees that the full reduced weight equals to a product of the reduced weights of all trajectories.
Fig. \ref{fig1} illustrates the rule of evaluation of reduced weights.

Consider the first trajectory from $x_1^0$ to $x_4$. It has three active intersections with trajectories starting at
$x_2^0$, $x_3^0$ and $x_4^0$. After each collision of two particles, only allowed parts of their trajectories survive so
that the hopping probabilities $v_1$, $v_2$, $v_3$, $v_4$ are ascribed to different pieces of trajectories as shown in
Fig. \ref{fig1}. Therefore the first trajectory has the first intersection with weight depending on $v_1$, the second one on
$v_2$ and the third on $v_3$.

The second trajectory from $x_2^0$ to $x_3$ has one passive and two active intersections so its total activity is $+1$.
The passive intersection has weight depending on $v_1$, which is compensated by the weight of the first active
intersection also depending on $v_1$. The weight of the second active intersection depends on $v_2$. Two other trajectories starting at $x_3^0$
and $x_4^0$ have two passive intersections each, depending on $v_1$, $v_2$ and $v_2$, $v_3$ correspondingly. We may
conclude that the reduced weight of each trajectory is defined by its number and its activity. The weight of $i$-th
$m$-active trajectory $(m>0)$ depends on $v_i$ and $m-1$ successive values $v_{i+1},\ldots,v_{i+m-1}$. The weight of
$i$-th passive trajectory of activity $-m$ depends on $v_{i-1},v_{i-2}\ldots,v_{i-m}$. Then we may write the resulting
expressions for the reduced weights in the operator form:

\begin{equation}
F_{i,m}\left(x_j-x_i^0,T\right)=\tilde{\mu}\left(
\prod_{k=0}^{m-1}\left(1-\frac{v_{i+k}}{1-v_{i+k}} \hat{a}_i \right)^{-1} \mathcal{T}_T(x_j|x_{i}^0)
\right),\;\; m>0,
\end{equation}
\begin{equation}
F_{i,m}\left(x_j-x_i^0,T\right)=\tilde{\mu}\left(
\prod_{k=m}^{-1}\left(1-\frac{v_{i+k}}{1-v_{i+k}} \hat{a}_i \right) \mathcal{T}_T(x_j|x_i^0)
\right),\;\; m<0
\end{equation}
and
\begin{equation}
F_{i,0}\left(x_j-x_i^0,T\right)=\tilde{\mu}\left( \mathcal{T}_T(x_j|x_i^0) \right).
\end{equation}
Analytical expressions for $F_{i,m}\left(x_j-x_i^0,T\right)$ are
\begin{equation}
F_{i,m}\left(x_j-x_i^0,T\right)=\oint\left(1+\frac{1}{z}\right)^T z^{x_j-x_i^0}
\prod_{k=0}^{m-1}\left(1-\frac{v_{i+k}}{1-v_{i+k}}z\right)^{-1}\frac{dz}{2\pi i z},\;\; m>0,
\end{equation}
\begin{equation}
F_{i,m}\left(x_j-x_i^0,T\right)=\oint\left(1+\frac{1}{z}\right)^T z^{x_j-x_i^0}
\prod_{k=m}^{-1}\left(1-\frac{v_{i+k}}{1-v_{i+k}}z\right)\frac{dz}{2\pi i z},\;\; m<0,
\end{equation}
and
\begin{equation}
F_{i,0}\left(x_j-x_i^0,T\right)=\oint\left(1+\frac{1}{z}\right)^T z^{x_j-x_i^0} \frac{dz}{2\pi iz}=\binom{T}{x_j-x_i^0}.
\end{equation}
Here the integration is over the circle around origin with vanishing radius.
Computing the weights for all permutations, one gets the $P\times P$ matrix {\bf M} with elements
\begin{equation}
M_{i,j}=F_{i,j-i}\left(x_j-x_i^0,T\right)
\end{equation}
and the conditional probability
\begin{equation}
P_T(\vec{x}|\vec{x}^0)=\mu({\mathbb S}_P\setminus{\mathbb U}_P)=Q \times \tilde{\mu}({\mathbb S}_P\setminus{\mathbb U}_P),
\end{equation}
where the weight factor $Q$ generalizes Eqs. (\ref{QPF}) and (\ref{Q2}):
\begin{equation}
Q=\prod_{i=1}^{P}v_i^{x_i-x_i^0}(1-v_i)^{T-x_i+x_i^0}.
\label{QP}
\end{equation}
Finally, the explicit expression for $P_T(\vec{x}|\vec{x}^0)$ generalizing Eq. \ref{detP} is
\begin{equation}
P_T(\vec{x}|\vec{x}^0)=\prod_{i=1}^{P}v_i^{x_i-x_i^0}(1-v_i)^{T-x_i+x_i^0} \det {\bf M}.
\label{PTZ}
\end{equation}
In the continuous time limit ($T\to\infty$, $v_i\to 0$, $0 < v_i T < \infty$, $v_i T \sim t\in\mathbb{R}^+$,
$i=1,2\ldots P$), we come to the result obtained by R\'{a}kos and Sch\"{u}tz \cite{RakosSchutz}.

To extend the derivation to the ring of length $L$, we map the trajectories wrapping the cylinder $\Lambda$ on the
infinite integer strip $\Omega = [-\infty,\infty]\times [0,T]$ introducing extended coordinates $X=x+nL$, $n$ integer, for
equivalent points. The initial coordinates of particles are $\vec{x}^0=(x_1^0,x_2^0,\ldots, x_P^0)$, $0\leq
x_1^0<x_2^0<\ldots < x_P^0 \leq L-1$. The positions of particles on the ring at time $T$ are given by ordered coordinates
$\vec{x}=(x_1,x_2,\ldots,x_P)$, $0\leq x_1<x_2<\ldots < x_P\leq L-1$ and cyclic permutation of indexes of particles
$\vec{\alpha}=(\alpha_1,\alpha_2,\ldots, \alpha_P)$. The ring coordinate of $i$-th particle $(i=1,2,\ldots,P)$ at
time $T$ is $x_{\alpha_i}$. Let $i$-th particle intersects the bond $[L-1,0]$ $n_{\alpha_i}^*$ times during time T.
Then the number of steps it advanced is
\begin{equation}
N_i=X_{\alpha_i}-x_i^0,
\end{equation}
where $X_{\alpha_i}=x_{\alpha_i}+n_{\alpha_i}^*L$ is the extended coordinate of $i$-th particle on the strip.
Since the particles cannot overtake one another, we have for any cyclic permutation $\vec{\alpha}$:
\begin{equation}
X_{\alpha_1} < X_{\alpha_2} < \ldots < X_{\alpha_P} < X_{\alpha_1} + L.
\label{overtake}
\end{equation}
Clearly, these inequalities impose restrictions on rotation numbers $\vec{n}^*=(n_1^*,n_2^*,\ldots,n_P^*)$.
Particularly, the set of independent variables describing unambiguously the state of system are the
ordered ring coordinates $\vec{x}$ and the total current through bond $[L-1,0]$
\begin{equation}
J=\sum_{i=1}^{P} n^*_i.
\end{equation}
Our aim is to find the conditional probability to find the particles in the ring positions for a fixed $J$ at time $T$ if
the particles start from initial positions $\vec{x}^0$.

\begin{figure}[h!]
\includegraphics[width=100mm]{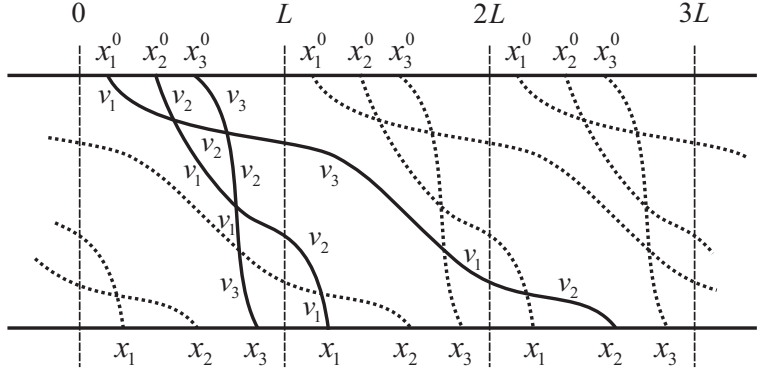}
\caption{\label{fig2} The configuration of three trajectories on the ring.}
\end{figure}

Consider a set of configurations of $P$ free trajectories on the strip $\Omega$ together with their copies placed
periodically with period $L$ in horizontal direction. We obtain the set of configurations
\begin{equation}
\mathbb{S}_P=
\mathcal{T}_T( x_{\alpha_1} + n_{\alpha_1}^* L |x_1^0 )\otimes
\mathcal{T}_T( x_{\alpha_2} + n_{\alpha_2}^* L |x_2^0 )\otimes\ldots\otimes
\mathcal{T}_T( x_{\alpha_P} + n_{\alpha_P}^* L |x_P^0 ).
\end{equation}
repeated on the strip infinitely many times.
As above, a configuration is allowed if there are no collision points among trajectories.
For an allowed configuration, there is at least one possibility to draw it without collisions.
To eliminate the unallowed set $\mathbb{U}_P$ of free configurations, we introduce a new auxiliary set.
In the case of periodical boundary conditions, this set includes not only the permutations of numbers of particles
$\vec{\pi}$, but also all possible numbers of rotations $\vec{n}=(n_1,n_2,\ldots,n_P)$:
\begin{equation}
\mathbb{A}_{\pi(1,2,\ldots,P),\;\vec{n}}=
\mathcal{T}_T( x_{\pi(1)} + n_{\pi(1)} L |x_1^0)\otimes
\mathcal{T}_T( x_{\pi(2)} + n_{\pi(2)} L |x_2^0)\otimes\ldots\otimes
\mathcal{T}_T( x_{\pi(P)} + n_{\pi(P)} L |x_P^0)
\end{equation}
The numbers of rotations are arbitrary but yield the condition
\begin{equation}
J=\sum_{i=1}^{P}n_i,
\label{J}
\end{equation}
because the permutation of end points in any pair of trajectories does not change the sum of rotation numbers.
The weights of configurations $\mathbb{S}_P$ and $\mathbb{A}_{\pi(1,2,\ldots,P),\;\vec{n}}$ as well as their reduced weights can be
defined in a way quite similar to that for the infinite lattice (Fig. \ref{fig2}).
The difference between the infinite lattice and the ring is a possibility for given
particle to overtake another one several times. For instance, the trajectory $x_1^0
\rightarrow x_2$ in Fig. \ref{fig2} has two active intersections with both trajectories
from $x_2^0$ to $x_1$ and from $x_3^0$ to $x_3$. So its total activity equals $m=4$. The
trajectory $x_3^0 \rightarrow x_3$ has three passive intersections and its activity is $m=-3$.
The trajectory $x_2^0 \rightarrow x_1$ has two passive and one active intersection, so its activity is $m=-1$.
In general, the activity of trajectory $x_i^0 \rightarrow x_j$ is given by
\begin{equation}
s_{ij}=j-i+P n_j-\sum_{k=1}^{P}n_k.
\end{equation}
The weights of different pieces of the trajectory starting from $x_i^0$ vary from $v_i$ to the next values
$v_{i+1}$ or $v_{i-1}$ for each active or passive intersection. If $i=P$, the next value
$v_{i+1}=v_{1}$ and if $i=1$, the value $v_{i-1}=v_{P}$.
Generally, the indexes of hopping probabilities coincide by modulo $P$:
\begin{equation}
v_{i + k P} \equiv v_i, \; k = \pm 1, \pm 2, \ldots
\end{equation}
The sign of a configuration is defined by the sign of permutation of end points.
Using these rules, we derive the reduced weight of allowed trajectories in the form
\begin{equation}
\tilde{\mu}({\mathbb S}_P\setminus{\mathbb U}_P) = \tilde{\mu}(\mathbb{S}_P) +
{\sum_{\vec{n}}}'
\sum_{\pi \neq \vec{\alpha}}
(-1)^{\sum\limits_{i<j}|n_i-n_j|}sgn(\pi)\,\tilde{\mu}(\mathbb{A}_{\pi(12\ldots P),\vec{n}}),
\end{equation}
where the first summation is over all integers $\{n_i \in \mathbb{Z}; i=1,2,\ldots, P\}$ that satisfy the relation (\ref{J}).
The second sum is over all permutations $\pi$ of particles except initial permutation $\vec{\alpha}$.
This expression can be written as a sum of determinants.
Introducing the matrix $M$
\begin{equation}
M_{i,j}=F_{i,s_{ij}}\left(x_j+n_j L-x_i^0,T\right),
\end{equation}
we obtain
\begin{equation}
\tilde{\mu}({\mathbb S}_P\setminus{\mathbb U}_P)=
{\sum_{\vec{n}}}'(-1)^{\sum\limits_{i<j}|n_i-n_j|} \det {\bf M}.
\end{equation}
The transition probability for fixed $J$ is defined by identity
\begin{equation}
P_T(\vec{x},J|\vec{x}^0)=
\mu({\mathbb S}_P\setminus{\mathbb U}_P)=
Q \times \tilde{\mu}({\mathbb S}_P\setminus{\mathbb U}_P),
\end{equation}
with the factor
\begin{equation}
Q=\prod_{i=1}^{P}v_i^{N_i}(1-v_i)^{T-N_i}.
\end{equation}
Finally, we find
\begin{equation}
P_T(\vec{x},J|\vec{x}^0)=\prod_{i=1}^{P}v_i^{N_i}(1-v_i)^{T-N_i}
{\sum_{\vec{n}}}'(-1)^{\sum\limits_{i<j}|n_i-n_j|} \det {\bf M}.
\label{PTJ}
\end{equation}
Being calculated explicitly, Eq. \ref{PTJ} is a polynomial in $v_1,\ldots,v_P$. For small $L$,$P$,$T$, it can be
found directly from consideration of allowed configurations of trajectories and compared with results of evaluation
of Eq. \ref{PTJ}. For example, in the case $L=5$, $P=3$, $T=4$, $\vec{x}^0=(0,1,2)$, $\vec{x}=(0,1,2)$ and
various $J$, we get
\begin{equation}
P_T(\vec{x},J=0|\vec{x}^0) = \left( 1 - v_3 \right)^4,
\end{equation}
\begin{eqnarray}
P_T(\vec{x},J=1|\vec{x}^0) = &&v_1 v_2 v_3^3 \left(1-v_2\right)^2
\left(  v_1^3 v_2 +v_1^3v_3 -2v_1^3 -7v_1^2v_2 + \right. \\
&&\left. 2v_1^2v_2v_3 -7v_1^2v_3 +12v_1^2 +18v_1v_2 -8v_1v_2v_3 + \right. \nonumber\\
&&\left. 18v_1v_3 -28v_1 -22v_2 +12v_2v_3 -22v_3 +32 \nonumber \right),
\end{eqnarray}
\begin{equation}
P_T(\vec{x},J=2|\vec{x}^0) = 6 \left( 1 - v_1 \right)^2 v_1^2 v_2^4 v_3^4.
\end{equation}
For $J>2$, the probability $P_T(\vec{x},J|\vec{x}^0)$ vanishes.

Summation over $J$ gives the transition probability from an arbitrary initial ring coordinates $\vec{x_0}$ to final
coordinates $\vec{x}$:
\begin{equation}
P_T(\vec{x}|\vec{x}^0)=\sum\limits_{J=0}^{\infty}P_T(\vec{x},J|\vec{x}^0).
\end{equation}
In the specific case, when the hopping probabilities $v_i$ are not particle-dependent, we can write $Q$ in the form
\begin{equation}
Q=v^{\sum\limits_{i=1}^{P}(x_i-x_i^0+n_i^*L)}(1-v)^{T-\sum\limits_{i=1}^{P}(x_i-x_i^0+n_i^*L)}.
\end{equation}
The conservation condition
\begin{equation}
J=\sum_{i=1}^{P}n_i^*=\sum_{i=1}^{P}n_i
\end{equation}
makes it possible to change all variables $n_i^*$ by $n_i$ and evaluate the sum over $J$. After the summation, we recover
the known result for equal hopping probabilities \cite{PriezCondmat,PriezPRL}:
\begin{equation}
P_T(\vec{x}|\vec{x}^0)=
\sum_{n_1=-\infty}^{+\infty} \ldots \sum_{n_P=-\infty}^{+\infty}
(-1)^{\sum\limits_{i<j}|n_i-n_j|}
\prod_{i=1}^{P}v^{x_i-x_i^0+n_i L}(1-v)^{T-x_i+x_i^0+n_i L}
\det {\bf M}.
\label{genfun}
\end{equation}
For considered values of parameters, it gives
\begin{equation}
P_T(\vec{x}|\vec{x}^0)=\left(1 - v\right)^2
\left( 6v^{10} + 4v^9 - 24v^8 + 60v^7 - 72v^6 + 32v^5 + v^2 - 2v + 1 \right).
\end{equation}
\section{Discussion}

We have calculated exactly the non-stationary conditional probability for the TASEP with particle-dependent
hopping probabilities on a ring. The steady state of this model has been studied by Evans \cite{Evans}, who found that
the mean velocity $v_{st}$ in the thermodynamic limit ($T\to\infty$, $L\to\infty$, $P\to\infty$, $P/L=\rho$)
is determined from equation
\begin{equation}
1 - \rho = v_{st} \frac{1}{L} \sum_{ i=1 }^{P} \frac{ 1 - v_i }{ v_i - v_{st} }
\end{equation}
Obviously, $v_{st}$ does not depend on the permutation of velocities $\vec{v}=(v_1,v_2,\ldots,v_P)$.
In the low-density phase the fast particles are sticking to slower ones and form a number of jams just behind slow particles.
The slow particles are mainly free, i.e. almost all of them have empty neighboring sites in direction of motion.
In the thermodynamic limit, the jam corresponding the slowest particle dominates over other clusters.
This behaviour has a strong analogy with Bose condensation and the steady-state velocity of a particle is equivalent to the
fugacity of an ideal Bose gas \cite{Evans}.
For high densities, the slowest particle is not free, the jam dissolves and a congested phase appears.

The non-stationary state of the TASEP considered in this paper differs from the stationary case in several respects.
First, the non-stationary mean velocity depends not only on absolute values of particle velocities, but on their order too.
Second, a time dependence of the total number of particles in the jam is an important characteristic of the traffic
\cite{Schreck1,Schreck2}.
Generally, the exact non-stationary solution of the TASEP on a ring may describe how the phase transition arises in
thermodynamic limit.
A next possible step in this study would be the TASEP with parallel dynamics, which answer the question whether distinct
updating schemes produce different collective behaviour.

\section*{Acknowledgments}
One of us (V.B.P.) is grateful to G.M. Sch\"{u}tz for useful and stimulating discussion.
This work was supported by RFBR grant No 06-01-00191a.

\end{document}